\documentclass[aps,prb,twocolumn,a4paper,showpacs,eqsecnum,floatfix,fleqn]{revtex4}
\usepackage{amsfonts}
\usepackage{amsmath}
\usepackage{bm}
\usepackage{natbib}
\usepackage{graphicx}
\usepackage{array}
\usepackage{dcolumn}

\begin{document}

\title{The energy-scale-dependent Composite Operator Method \protect\\ for the single-impurity Anderson model}
\author{Adolfo \surname{Avella}}
\email[E-mail address: ]{avella@sa.infn.it}
\affiliation{Dipartimento di Fisica ``E.R. Caianiello'' - Unit\`a di Ricerca INFM di Salerno\\
Universit\`a degli Studi di Salerno, I-84081 Baronissi (SA),
Italy}
\author{Roland \surname{Hayn}}
\affiliation{Institut f\"ur Festk\"orper- und Werkstofforschung
(IFS) Dresden, D-01171 Dresden, Germany}
\author{Ferdinando \surname{Mancini}}
\affiliation{Dipartimento di Fisica ``E.R. Caianiello'' - Unit\`a di Ricerca INFM di Salerno\\
Universit\`a degli Studi di Salerno, I-84081 Baronissi (SA),
Italy}

\date{\today}

\begin{abstract}
The recently developed energy-scale-dependent Composite Operator
Method is applied to the single-impurity Anderson model. A fully
self-consistent solution is given and analyzed. At very low
temperatures, the density of states presents, on the top of the
high-energy background, a Kondo-like peak whose parameter
dependence is discussed in detail. The proposed method reproduces
the exact results known in the literature with very low numerical
effort and it is applicable for arbitrary values of the external
parameters.
\end{abstract}


\maketitle

\section{Introduction}

The theoretical description of strongly correlated electron
systems, like transition-metal oxides\cite{Imada:98} and heavy
fermion compounds\cite{Gunnarsson:87}, is of high actual interest
and effective analytical methods to study them are looked for. A
general problem consists in connecting the high- and low- energy
scales in a proper way. Methods based on the use of the equations
of motion for the Green's functions (e.g., the projection
methods\cite{Mori,Rowe:68,Roth:69,Nolting:72,Tserkovnikov:81,Nolting:89,Plakida:89,Fedro:92,Fulde:95,Manciniredux,Matsumoto}),
usually give a rather reliable description of the high-energy
features of strongly correlated systems, but, quite often, the
overall solution obtained by their application does not reproduce
the low-energy physics accurately enough. Other techniques, like
the slave-boson approximation\cite{Barnes:76}, provide a correct
picture at low energies but fail at higher ones. However, there is
an emerging consensus\cite{Georges:96} that in strongly correlated
electronic systems we should search for a correct description of
both energy scales at once. The anomalous behaviors shown by these
systems are caused and/or influenced by both the broad incoherent
spectral features far away from the Fermi level and the more
dispersive quasi-particle bands close to it.
Recently\cite{Villani:00}, it was shown how the Composite Operator
Method\cite{Manciniredux,Matsumoto} (COM) could be used to resolve
coherent low-energy features in a proper way by solving the SU(N)
Kondo model.

The Kondo model should be considered just as a first step. It
contains no charge degrees of freedom of the strongly correlated
impurity level, and it is usually not sufficient for a realistic
description of 4$f$-electron spectra. For this latter purpose the
single-impurity Anderson model is much better
suited\cite{Gunnarsson:87}. Furthermore, the solution of the
Anderson model is a building block within the Dynamical Mean Field
Theory\cite{Georges:96} (DMFT) algorithm, which has favored large
progresses in the comprehension of the Mott metal-insulator
transition phenomenon. The single-impurity Anderson model is well
and widely studied\cite{Hewson:97}, but the reliable methods are
rather involved (e.g., quantum Monte Carlo method\cite{Hirsch:86},
Bethe ansatz\cite{Wiegmann:83}, numerical renormalization
group\cite{Wilson:75} and non-crossing
approximation\cite{Hewson:97}) and require a quite huge
computational effort. This is certainly an obstacle to use them
within the DMFT algorithm or to interpret the realistic spectra of
rare-earth compounds. According to this, we present here a simple
analytical method to solve the single-impurity Anderson model,
which is capable to reproduce both the high- and the low- energy
features, as known after the exact solution\cite{Hewson:97}, in a
reasonable way and, practically, without requiring any
computational effort at all.

\section{Hamiltonian \protect\\ and equations of motion}

The single-impurity Anderson model is defined by the Hamiltonian
\begin{multline}
H = \sum_{{\bf k},\sigma}\varepsilon_{\bf k} c_{{\bf
k}\sigma}^\dagger c_{{\bf k}\sigma} + \sum_\sigma \varepsilon_f
f_\sigma^\dagger f_\sigma + U n_{f \downarrow} n_{f \uparrow} \\
+ \frac{V}{\sqrt{N}} \sum_{{\bf k},\sigma} (c_{{\bf
k}\sigma}^\dagger f_\sigma + f_\sigma^\dagger c_{{\bf k},\sigma})
\; , \label{e01}
\end{multline}
where $c_{{\bf k}\sigma}$ represents the electrons in the valence
band ($\varepsilon_{\bf k}$) and $f_\sigma$ those at the impurity
level ($\varepsilon_f$); $n_{f\sigma}$ is the density charge
operator for the $f$-electrons of spin $\sigma$, $N$ is the number
of sites, $U$ is the Coulomb repulsion at the impurity level and
$V$ is the strength of the hybridization between the valence band
and the impurity level. For the sake of simplicity, we restrict
ourselves to a simple two (spin) degenerate impurity level
($\sigma=\pm 1$, $\bar \sigma = - \sigma$).

The first step within any method based on the projection technique
consists in individuating an appropriate set of basis operators.
In fact, for strongly correlated systems the original electronic
operators are just the wrong place where to start any approximate
treatment\cite{Mancini:00}. The electrons completely lose their
identities owing to the strong interactions and other complex
excitations appear. These latter are the only effective
quasi-particles present in the systems and only in terms of them
any description of the dynamics should be attempted. The basis
(composite) operators should be then chosen in order to describe
such excitations. The choice is not easy, but some recipes could
be given\cite{recipe}. In the present case one should have as
minimal requirement an appropriate description of the free
propagation of $c$-electrons and the atomic dynamics of
$f$-electrons. The latter have two possible excitations
corresponding to the transitions from $|0\rangle_f$ to
$|\sigma\rangle_f$ and from $|\sigma\rangle_f$ to
$|\uparrow\downarrow\rangle_f$. That leads to the first three
basis operators
\begin{equation}
\begin{split} \label{b1}
&\psi_{1\sigma} = c_{0\sigma} =\frac{1}{\sqrt{N}}
\sum_{\bf k} c_{{\bf k}\sigma} \\
&\psi_{2\sigma} = \xi_{\sigma} = (1 - n_{f\bar \sigma}) f_{\sigma} \\
&\psi_{3\sigma} = \eta_{\sigma} = n_{f\bar \sigma} f_{\sigma} \; .
\end{split}
\end{equation}
It is worth noting that $\psi_{2\sigma} = \xi_{\sigma}=X^{0
\sigma}$ and $\psi_{3\sigma} = \eta_{\sigma} = \sigma X^{\bar
\sigma 2}$ are the Hubbard operators describing the lower- and the
upper- Hubbard subbands.

These three operators give a very good description of the
high-energy features, but are certainly insufficient to describe
the low-energy ones and the mixing between the two regimes. Then,
in order to overcome such limitations, we write down the equations
of motion for the operators (\ref{b1})
\begin{equation}
\begin{split} \label{Eqeq}
& \mathrm{i}\partial_t c_{\bf k \sigma} = [ c_{{\bf k}\sigma}, H ]
= \varepsilon_{\bf k} c_{{\bf k}\sigma}
+ \frac{V}{\sqrt{N}} ( \xi_{\sigma} + \eta_{\sigma} ) \\
& \mathrm{i}\partial_t \xi_{\sigma} = [ \xi_{\sigma} , H ] =
\varepsilon_f \xi_{\sigma} + \frac{V}{2}
c_{0 \sigma} + V \pi_{\sigma} \\
& \mathrm{i}\partial_t \eta_{\sigma} = [ \eta_{\sigma} , H ] =
(\varepsilon_f + U) \eta_{\sigma} + \frac{V}{2} c_{0 \sigma} - V
\pi_{\sigma} \; ,
\end{split}
\end{equation}
and account for the appearance of a new operator, namely the
fluctuation field
\begin{equation}
\pi_{\sigma}= \psi_{4\sigma}+\psi_{5\sigma}+\psi_{6\sigma} \; ,
\end{equation}
with
\begin{equation}
\begin{split} \label{b2}
&\psi_{4\sigma} = \frac{1}{2} (1-n_f ) c_{0\sigma} \\
&\psi_{5\sigma} = \sigma c_{0 \sigma} S^z + c_{0 \bar{\sigma}}
S^{\bar{\sigma}} \\
&\psi_{6\sigma} = c_{0 \bar{\sigma}}^{\dagger} f_{\bar{\sigma}}
f_{\sigma} \; ,
\end{split}
\end{equation}
where $n_f=n_{f \uparrow} + n_{f \downarrow} $,
$S^z = (n_{f \uparrow} - n_{f \downarrow} ) / 2$ and
$S^{\sigma}=f_{\sigma}^{\dagger} f_{\bar{\sigma}}$. The
fluctuation field $\pi$ describes the coupling of the valence band
to density-, spin- and pair- impurity fluctuations and opens the
possibility to study the low-energy dynamics connected with them.
According to this, we have decided to include also the operators
(\ref{b2}) into the basis and study the system in terms of this
set of six basic excitations. Such basis cannot be considered
complete according to the fact that an infinite degree of freedom
system possesses an infinite number of operator basis. Then, it is
obvious that the basis is neither unique. In particular, it has
been already mentioned above that many recipes can be used in
order to construct an operatorial basis, each according to the
features of interest. However, as we will see from the obtained
results, this basis seems sufficient to catch the main features of
the dynamics of this system.

In order to be more compact, we rewrite the six basis operators in
spinorial notation
\begin{equation}\label{basis}
\begin{array}{ll}
\psi_1 = c_{0} = \frac{1}{\sqrt{N}} \sum_{\bf k} c_{{\bf k}} \quad
& \quad \psi_4 = \frac12(1-n_f) c_{0} \\
\psi_2 = \xi= (1 - n_f) f  \quad &  \quad \psi_5 =
\frac12\vec{\sigma} \circ \vec{n}_f \cdot c_{0} \\
\psi_3 = \eta = n_f f  \quad  & \quad \psi_6 = c_{0}^\dagger \cdot
\xi \otimes \eta \; .
\end{array}
\end{equation}
Here $\vec{n}_f = f^{\dagger} \cdot \vec{\sigma} \cdot f$,
$\vec{\sigma}$ are the Pauli matrices, $\cdot$ denotes the scalar
product in spin space, $\circ$ the scalar product in direct space
and $\otimes$ the tensor product. This notation will be used
hereafter. It should be noted that the exact expressions of the
basis operators have been chosen such that they transform under
the particle-hole transformation ($c_0 \to c^\dagger_0$ and $f \to
f^\dagger$) into themselves or into another basis operator:
$\psi_1 \to \psi^\dagger_1$, $\psi_2 \to \psi^\dagger_3$, $\psi_3
\to \psi^\dagger_2$ and $\psi_n \to -\psi^\dagger_n$ for
$n=4,5,6$.

As next step, we derive the equations of motion for the retarded
Green's functions (GF)
\begin{eqnarray}
G_{nm}(\omega) &=& \mathcal{F}\langle  \mathcal{R} \left[
\psi_n(t)
\psi_m^\dagger(t')  \right] \rangle  \nonumber \\
&=& \mathcal{F} \left[ \theta(t-t') \langle \{ \psi_n(t),
\psi_m^\dagger(t') \}\rangle \right]
\end{eqnarray}
where $\mathcal{F}$ stays for the Fourier transform and
$\mathcal{R}$ for the retarded time ordering operator. The
inhomogeneous term in the equations of motion for the Green's
functions $G_{nm}$ is constituted by the normalization matrix (its
expression in terms of basic field correlators is given in
Eq.~(\ref{e5}) of App.~A)
\begin{equation}
I_{nm}=\langle \{\psi_n(t),\psi_m^\dagger(t)\} \rangle \; ,
\end{equation}
where $\{-,-\}$ stays for the anticommutator and $\langle -
\rangle$ for the thermal average. The normalization matrix is not
only fundamental to derive the Green's functions but provides also
important information about the total spectral weights of the
fields. Its matrix elements depend on the expectation values
$C_{ij}=\langle \psi_i \psi_j^{\dagger} \rangle$ (correlation
matrix). It contains for instance the average charge density at
the impurity level $\langle n_f \rangle = 2 (1 - C_{22} -
C_{33})$, and the double occupancy $D_f = \langle n_{f \uparrow}
n_{f \downarrow} \rangle = 1 - C_{22} - 2 C_{33}$. For further use
we define also the matrix elements with the fluctuation field,
i.e., $I_{n\pi}=\langle\{\psi_n(t),\pi^\dagger(t)\} \rangle$.
Then, after the equations of motion (\ref{Eqeq}), we have the
following expressions for the Green's functions
\begin{eqnarray}
\label{e6} G_{11} &=& \Gamma_0 + V^2 \Gamma_0^2 \left( G_{22} + 2
G_{23} +
G_{33} \right) \nonumber \\
G_{22} &=& I_{22} \frac{\Gamma_{+}}{F} + \frac{B_+^2}{F^2}
G_{\pi \pi} + \frac{B_+(C_++x_+C_-)}{F^2} \nonumber \\
G_{33} &=& I_{33} \frac{\Gamma_{-}}{F} + \frac{B_-^2}{F^2}
G_{\pi\pi} - \frac{B_-(C_-+x_-C_+)}{F^2}  \\
G_{23} &=& x_- G_{22} - \frac{V \Gamma_{-}}{F} \left( B_+ G_{\pi
\pi} + C_++x_+C_- \right) \nonumber
\end{eqnarray}
where we introduced the following abbreviations
\begin{equation}
\begin{array}{ll}
\Gamma_0 = \frac{1}{N} \sum_{\bf k }
\frac{1}{\omega-\varepsilon_{\bf k}} & \Gamma_{s} =
\frac{1}{\omega-\varepsilon_s -V^2 \Gamma_0 / 2}
\\
x_s = \frac{1}{2} V^2 \Gamma_0 \Gamma_{s}  &
B_s = V \Gamma_{s} - V x_s \Gamma_{\bar s} \\
C_+ = (I_{2\pi} + \frac{1}{2} V \Gamma_0 I_{1 \pi} ) \Gamma_{+} &
C_- = (I_{3\pi} + \frac{1}{2} V \Gamma_0 I_{1 \pi} ) \Gamma_{-} \\
F = 1 - x_+ x_-  & \\
\end{array}
\end{equation}
with $s=\pm$, $\bar s = -s$, $\varepsilon_+=\varepsilon_f$, and
$\varepsilon_-=\varepsilon_f+U$, to shorten the notation. It is
worth noticing that $\Gamma_0(\omega)$ is just the free ($V=0$)
propagator for $c_0$ operator. We see from (\ref{e6}) that we can
calculate the Green's function for the valence electrons $G_{11}$
and the $f$-impurity $G_{ff}=G_{22}+2G_{23}+G_{33}$ once we know
the Green's function of the fluctuation field $G_{\pi \pi}$.

\section{High- and low- energy scales}

In order to resolve the low-energy features embedded in the
high-energy background, following the idea given in
Ref.~\onlinecite{Villani:00}, which is based on the
well-established physical assumption that at low energies we have
a quasi-particle theory\cite{Hewson:97} as also derived by the
slave-boson theory\cite{Barnes:76}, we split the dynamics of the
fluctuation field into an high- and a low- energy part. As the
essence of the Kondo effect consists in the coupling of the
valence band to the spin fluctuations at the impurity level, we
split only $\psi_5$ as
\begin{equation}\label{fieldsplit}
\psi_{5} =\psi_{5}^H + \psi_{5}^L
\end{equation}
and assume instead that the charge ($\psi_{4}$) and the pair
($\psi_{6}$) terms in the fluctuation field are sufficiently well
represented by their high-energy parts only ($\psi_{4}=\psi^H_{4}$
and $\psi_{6}=\psi^H_{6}$). In practice, we assume that the
low-energy field $\psi_5^L$ spans a different energy sector of the
Hilbert space with respect to $\psi_5^H$, $\psi_4$, $\psi_6$, and
describes a coherent quasi-particle at very low energies; energies
that are much smaller than any other defined in the Hamiltonian.
According to this, we make the following ansatz
\begin{equation}
\mathrm{i}\partial_t \psi_5^L = [ \psi_5^L , H ] = \kappa_1 c_{0}
+ \kappa_2 \xi + \kappa_3 \eta \label{Eq9}
\end{equation}
The coefficients $\kappa_i$ ($i=1,2,3$) are determined by
projecting onto the basis (\ref{basis})
\begin{eqnarray}
\kappa_1 &=& V (I_{25}^L + I_{35}^L ) \nonumber \\
\kappa_2 &=& \varepsilon_f I_{25}^L / I_{22}  + V I_{55}^L / ( 2
I_{22} )  \\
\kappa_3 &=& (\varepsilon_f + U) I_{35}^L / I_{33}  - V I_{55}^L /
( 2 I_{33} )  \nonumber
\end{eqnarray}
with
\begin{equation}
I_{5i}^L = \langle \{ \psi_5^L , \psi_i^\dagger \} \rangle \qquad
\end{equation}

After Eq.~\ref{fieldsplit} and the above reported consequent
reasoning, we split the still unknown Green's function $G_{\pi
\pi}=\mathcal{F}\langle \mathcal{R} \left[ \pi (t) \pi (t')
\right] \rangle$ into a low-energy component $G_{\pi
\pi}^L=\mathcal{F}\langle \mathcal{R} \left[ \pi^L(t) \pi^L(t')
\right] \rangle=\mathcal{F}\langle \mathcal{R} \left[ \psi_5^L(t)
\psi_5^L(t') \right] \rangle=G_{55}^L$ and in a high-energy one
$G_{\pi \pi}^H=\mathcal{F}\langle \mathcal{R} \left[ \pi^H(t)
\pi^H(t') \right] \rangle=\sum_{n,m=4}^6 \mathcal{F}\langle
\mathcal{R} \left[ \psi_n^H(t) \psi_m^H(t') \right] \rangle$. We
neglect the cross-term $\mathcal{F}\langle \mathcal{R} \left[
\pi^L(t) \pi^H(t') \right] \rangle$ according to our previous
assumption of no-overlap of the energy sectors spanned by
$\psi_5^L$ and all other fields involved. According to
Eq.~(\ref{Eq9}), $G_{55}^L$ obeys the following equation of motion
\begin{equation}
\omega G_{55}^L = I_{55}^L + \kappa_1 G_{15}^L +\kappa_2 G_{25}^L
+ \kappa_3 G_{35}^L
\end{equation}
where $G_{i5}^L=\mathcal{F}\langle \mathcal{R} \left[ \psi_i(t)
\psi_5^L(t')  \right] \rangle$. Then, for the sake of consistency,
we approximate $G_{i5}^L$ (see appendix for complete expressions)
only by those components that are explicitly proportional to
$G_{55}^L$, i.e.,
\begin{equation}
\begin{split} \label{e14}
&G_{15}^L = V \Gamma_0 ( G_{25}^L + G_{35}^L )\\
&G_{25}^L = \frac{B_+}{F} G_{55}^L \\
&G_{35}^L = -\frac{B_-}{F} G_{55}^L
\end{split}
\end{equation}
and we finally obtain
\begin{equation}
\begin{split} \label{e15}
& G_{\pi \pi
}^L = G_{55}^L =\frac{I_{55}^L}{\omega-\Omega_0} \\
&\Omega_0 = \frac{(\kappa_1 V \Gamma_0 + \kappa_2 ) B_+ -
(\kappa_1 V \Gamma_0 + \kappa_3 ) B_- }{2 F} \; .
\end{split}
\end{equation}
It is worth noticing that, in determining the actual expressions
for $G_{i5}^L$, we have neglected both the cross-terms
$\mathcal{F}\langle \mathcal{R} \left[ \psi_i^H(t) \psi_5^L(t')
\right] \rangle$ with $i=4,5,6$, according to our previous
assumption of no-overlap of the energy sectors, and those terms
proportional to $\Gamma$ and $\Gamma_s$, according to the
well-defined high-energy character of these latter.

We are now left with the task of computing the high-energy
contribution $G_{\pi\pi}^H = \sum_{n,m=4}^6 G_{nm}^H$ where
$G_{nm}^H = \mathcal{F}\langle \mathcal{R} \left[ \psi_n^H(t)
\psi_m^H(t') \right] \rangle$. In order to accomplish this task,
we use the mode-coupling approximation\cite{Bosse:78}, also known
as self-consistent Born approximation. In practice, we neglect the
mixing terms among different bosonic modes (i.e., we take
$G_{45}^H=G_{46}^H=G_{56}^H=0$) and decouple the remaining
corresponding time-ordered propagators in terms of the charge
($S_0=\mathcal{F}\langle  \mathcal{T} \left[ n_f(t) n_f(t')
\right] \rangle$), spin ($S_z=\mathcal{F}\langle \mathcal{T}
\left[ S^z(t) S^z(t') \right] \rangle$) and pair
($S_p=\mathcal{F}\langle  \mathcal{T} \left[ f_{\downarrow}(t)
f_{\uparrow}(t) f^\dagger_{\uparrow}(t')
f^\dagger_{\downarrow}(t') \right] \rangle$) impurity time-ordered
propagators, and of the valence electron time-ordered propagator
($S_{11}$). For instance, we have
\begin{equation}
S_{55}^H (\omega) = 3\frac{\mathrm{i}}{2\pi}\int d\Omega
S_z(\omega-\Omega) S_{11}(\Omega)
\end{equation}
Actually, for energies as high as those we intend to describe in
the high-energy sector and as far as the impurity level is not too
deep inside the valence band (i.e., for $|\varepsilon_f|$ not too
small with respect to the bandwidth) we can safely take the atomic
limit for the spin, charge and pair impurity propagators. In the
high-energy regime, any other treatment would not substantially
affect our results (i.e., by taking the atomic limit we already
get an accuracy of the same order of the energy scale we are
computing). For instance, we have
\begin{equation}
S_z(\omega)=-\frac14 2\mathrm{i}\pi \left[\langle n_f
\rangle-2D_f\right]\delta(\omega)
\end{equation}
Then, in the symmetric case ($- 2 \varepsilon_f=U$, leading to
$\langle n_f \rangle = 1$), all the non-zero contributions are
proportional to $G_{11}$, namely
\begin{equation}
\begin{split} \label{hep}
&G_{44}^H = \frac{1}{4}(1-\langle n_f \rangle+2D_f) G_{11} \\
&G_{55}^H = \frac{3}{2}(\langle n_f \rangle/2-D_f)) G_{11} \\
&G_{66}^H = -D_f G_{11}
\end{split}
\end{equation}

Finally, we can write the following expression for the fluctuation
field Green's function $G_{\pi \pi}$, which take into account both
high- and low- energy contributions
\begin{equation}
G_{\pi \pi} = (\frac{1}{4} + \frac{\langle n_f \rangle}{2}-2D_f)
G_{11} + G_{55}^L \; .
\label{e16}
\end{equation}
For $\varepsilon_f$ outside the valence band we have numerically
checked that the influence of the double occupancy $D_f$ in
Eq.~(\ref{e16}) is very small and could be neglected. This latter
procedure has been adopted in the asymmetric case (as $U$ has
always been taken much larger than the bandwidth) with the result
that Eq.~(\ref{e16}), with $D_f=0$, also holds.

It is worth noting that the high-energy Green's function
$G_{55}^H$ exhausts only part of the spectral weight $I_{55}$, as
expected from an independent evaluation of normalization matrix,
\begin{equation}
\lim_{\omega\gg1} G_{55}^H = \frac34 \frac{\left(\langle n_f
\rangle -2D_f \right)}{\omega} \neq \frac{I_{55}}{\omega} =
\lim_{\omega\gg1} G_{55}
\end{equation}
which can be derived from (\ref{hep}) using $\lim_{\omega\gg1}
G_{11} = 1/\omega$. On the contrary, we have: $\lim_{\omega\gg1}
\omega G_{44}^H = I_{44} = \lim_{\omega\gg1} G_{44} $ and
$\lim_{\omega\gg1} \omega G_{66}^H = I_{66} = \lim_{\omega\gg1}
G_{66} $. This consideration is fundamental as explains why we are
allowed/forced to consider the possibility of a low-energy
contribution to the dynamics and why we can/have to concentrate on
the spin fluctuations.

\section{Self-consistency}

In order to finally compute the GF in a self-consistent way, one
has to take into account two aspects. First of all, we have to
determine the matrix elements of the normalization matrix by
calculating the expectation values
\begin{equation}
C_{ij}=\langle \psi_i \psi_j^{\dagger} \rangle = \int d \omega
[1-f_\mathrm{F}(\omega)] A_{ij} (\omega) \label{flt}
\end{equation}
where $A_{ij} (\omega)=-\frac{1}{\pi}\Im [G_{ij} (\omega)]$ and
$f_\mathrm{F}$ is the Fermi function. Second, a self-consistent
calculation of the low-energy parts of the spectral weights
$I_{25}^L$, $I_{35}^L$, and $I_{55}^L$ is needed in order to
determine $G_{55}^L$ (see Eq.\ (\ref{e14}). For that purpose we
start from the total spectral weights
$I_{25}=-3 C_{12}/2$, $I_{35}=-3 C_{13}/2$, and $I_{55}=\frac34
\left(\langle n_f \rangle -2D_f \right)+C_{15}$. Then, for a given
temperature $T$ and for given low-energy contributions $I^L$, one
can determine the total weights $I$, written schematically as
$I[T,I^L]$, by exploiting the well-known connection between the
correlation and the Green's function matrices (\ref{flt}). To
extract the weight connected with the low-energy part, we compare
the spectral weights calculated for $I^L=0$ (defining
$I^H[T]=I[T,I^L=0]$) with the complete expression. Then, the
required self-consistency condition can be written as
\begin{equation}
I^L[T]=I[T,I^L]-I^H[T]
\end{equation}
This procedure has demonstrated to be capable to grasp part of the
effects connected with the mixing between the high- and low-
energy sectors. To calculate $C_{12}$, $C_{13}$, and $C_{15}$ we
need also the corresponding Green's functions whose expressions
are given in the appendix for completeness.

\section{Numerical results}

The numerical calculations have been performed with a density of
states for the valence band electrons $c({\bf k})$ which is
constant and finite only between energies $-D$ and $D$. That is,
we have used a free ($V=0$) propagator $\Gamma_0(\omega)$ for
$c_0$ with the following well-known expression \cite{Mahan:90}
\begin{equation}
\Gamma_0 (\omega) = \frac{1}{2D} \ln \left|
\frac{D+\mu+\omega}{D-\mu-\omega} \right| - \frac{i \pi}{2D}
\theta(D-|\omega+\mu|)
\end{equation}
We have given the formula for an arbitrary value of the chemical
potential $\mu$. The expectation values at finite temperature $T$
have been calculated by means of the Matsubara formalism
\begin{equation}
C_{lm}= -2 T \Re [\sum_{n=0}^{\infty} (G_{lm}(i\omega_n)
-\frac{I_{lm}}{i\omega_n-\varepsilon})] + [1-f_\mathrm{F}
(\varepsilon)] I_{lm}
\end{equation}
where $\omega_n=(2n+1)\pi T$. The arbitrary small number
$0<\varepsilon \ll 1$ and the Fermi function $f_\mathrm{F}$ have
been introduced in order to ensure the correct behavior at high
frequencies. The zero-order Green's function along the imaginary
axis is given by
\begin{multline}
\Gamma_0 (i \omega) = \frac{1}{2D} \ln \sqrt{
\frac{(D+\mu)^2+\omega^2}{(D-\mu)^2+\omega^2}} \\
+ \frac{i}{2D} \left( \arctan \frac{\omega}{D+\mu} + \arctan
\frac{\omega}{D-\mu} - \pi \right)
\end{multline}

\begin{figure}[p!!]
\includegraphics[width=8cm,keepaspectratio=true]{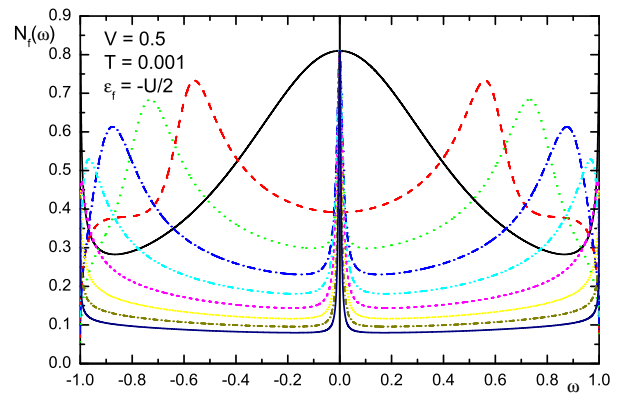}
\caption{Impurity density of states in the symmetric case for
different positions of the impurity level. The parameters are
$V=0.5$, $T=0.001$, $D=1$, and $U=0$, $1.2$, $1.6$, $2.0$,
$\ldots$, $4$.} \label{Fig1}
\end{figure}

\begin{figure}[p!!]
\includegraphics[width=8cm,keepaspectratio=true]{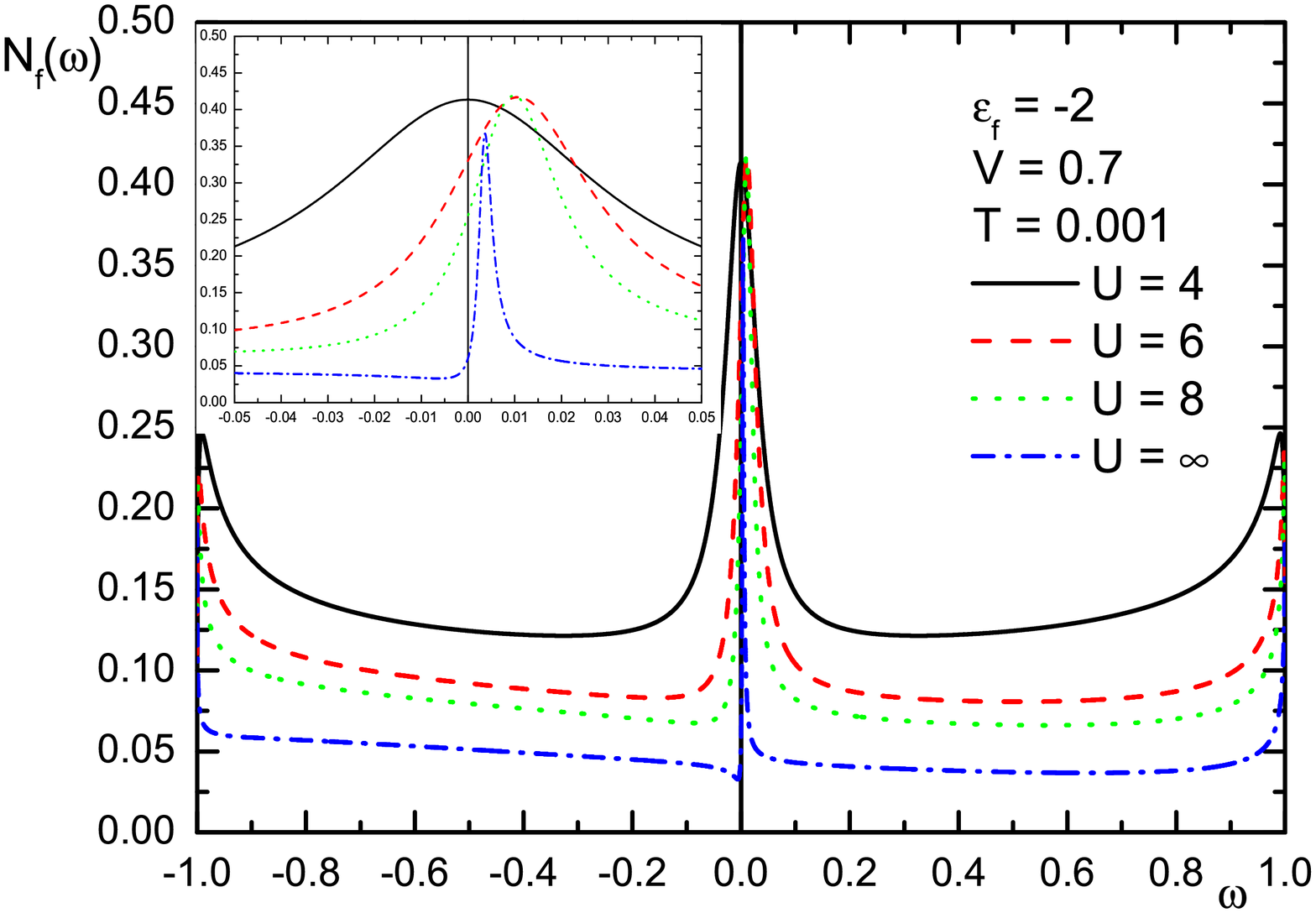}
\caption{Kondo peak in the asymmetric case with the parameters
$\varepsilon_f=-2$, $V=0.7$, $D=1$, $\mu=0$, and $T=0.001$. The
width of the Kondo peak diminishes with increasing Hubbard
correlation $U$ [$I_{55}^L=0.374$ for $U=4$ (symmetric case),
$0.287$ for $U=6$, $0.215$ for $U=8$ and $0.068$ for $U=\infty$].}
\label{Fig2}
\end{figure}

\begin{figure}[p!!]
\includegraphics[width=8cm,keepaspectratio=true]{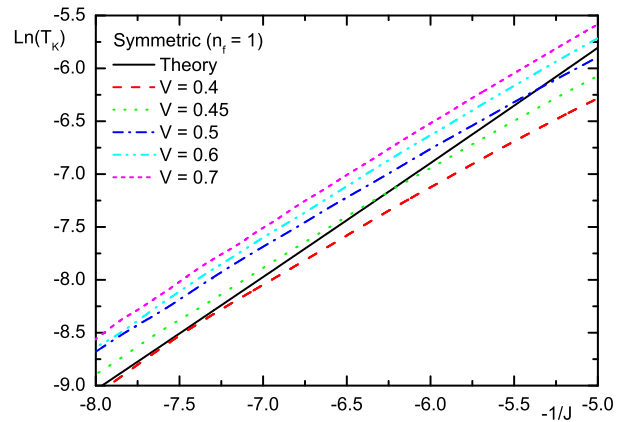}
\caption{Kondo-temperature in the symmetric case in dependence on
the exchange coupling $J$ compared with the formula given in the
text.} \label{Fig3}
\end{figure}

Let us start the discussion regarding our results with a short
briefing about the high-energy features. It is worth noticing that
our solution manage to describe the physics at the energy scale of
$U$ in a very effective way by explicitly taking into account the
excitations related to the Hubbard subbands. In Fig.~\ref{Fig1},
in fact, we can clearly see that the relevant features are
correctly reproduced: the splitting of the non-interacting band
into two subbands for finite values of $U$, a distance in energy
between the centers of mass of the two subbands almost equal to
$U$, spectral weights independent of bare $U$, but strongly
dependent on the ratio $V/U$. As regards the scale of energy $V$,
we use a basis rich enough to give the exact solution in absence
of $U$: second order ($V^2$) resonant behavior, spectral weights
satisfying the ordinary sum rules. Then, higher order processes
ruled by the competition of these two scales of energy (the
localizing $U$ and the dispersing $V$), and strictly dependent
only on the ratio $V/U$, are just those responsible for the
low-energy features we will discuss in the following paragraphs.

We can now move to the study of the low-energy features (whose
description is the actual goal of this manuscript) and, in
particular, to the analysis of the evolution of the Kondo peak by
moving the impurity level towards the valence band in the
symmetric case (see Fig.~\ref{Fig1}). The peak is widened for
decreasing values of $|\varepsilon_f|$ outside of the band. In
that case we have two bound-states at roughly $\pm \varepsilon_f$
which are not shown in Fig.~\ref{Fig1}. For $|\varepsilon_f|$
inside of the band we can observe a characteristic three peak
structure. Remarkably, the central peak height is not changed and
identical to the height for $U=0$, in agreement with the Fridel
sum rule. All these features are in agreement with the exact
behavior known from the numerical renormalization group
\cite{Hewson:97}.

Fig.~2 shows results in the asymmetric case. The Kondo peak is
shifted slightly above the Fermi level and its shape becomes
asymmetric. Furthermore, the width of the Kondo peak is decreased
in agreement with the reduction of the Kondo temperature. It is
worth noting that the Kondo peak also remains for $U \to \infty$.
If we increase the temperature, the Kondo peak vanishes by
diminishing its width (not shown). We have defined the Kondo
temperature as the one at which the parameter $I_{55}^L$ has a
change in the concavity when plotted as a function of the
temperature. The Kondo temperature becomes exponentially small for
small values of $J=V^2(1/|\varepsilon_f| + 1/|\varepsilon_f+U|)$
in agreement (see Fig. 3) with the formula\cite{Hewson:97} $T_K=D
\sqrt{2 J \rho_0} \exp{[-1/(2J\rho_0)]}$, where $\rho_0=1/(2D)$ is
the density of states of the unperturbed valence band at the Fermi
level. It is worth noting that this formula is correct only in the
symmetric case and for $J\ll D$. We have a strong dependence on
the value of the hybridization $V$ that shows a lack of
universality. Anyway, the linear behavior at small $J$ is
preserved for any value of $V$ showing that the exponential
dependence, which is not possible to obtain perturbatively, is
correctly described. We also studied the infinite bandwidth case,
but it does not give any qualitative change to the physical
picture described above.

In conclusion, a recently developed energy-scale-dependent
approach\cite{Villani:00}, which is capable to reproduce in a
reasonable way both high- and low- energy features of known exact
solutions of impurity models, has been extended to the case where
relevant charge fluctuations are present. The originating
procedure, the Composite Operator
Method\cite{Manciniredux,Matsumoto}, provides a fully
self-consistent solution where, on the top of a broad high-energy
background, a Kondo peak is present at low temperatures. The
parameter dependencies of the peak features and of the Kondo
temperature have been correctly reproduced, with respect to the
exact results known in the literature, with very low numerical
effort.

\acknowledgements{A.A. wishes to thank Dario Villani and Gabriel
Kotliar for many useful discussions on the subject and for all the
preliminary work done together.}

\appendix
\section{The normalization matrix}
The normalization matrix is found to be
\begin{widetext}
\begin{equation}
I=\left(
\begin{array}{cccccc}
1 & 0 & 0 & \frac12 \left(1- \langle n_f \rangle\right) & 0 & 0 \\
0 & 1 - \langle n_f \rangle / 2 & 0 &  -\frac12 C_{12} & -\frac32
C_{12} & C_{13} \\
0 & 0 & \langle n_f \rangle / 2 &   -\frac12 C_{13} & -\frac32
C_{13} & C_{12} \
\\
\frac12 \left(1- \langle n_f \rangle\right)  &  -\frac12 C_{12} &
-\frac12 C_{13} & \frac14 \left(1 - \langle n_f \rangle\right) +
\frac12 D_f & 0 &  \frac12 C_{16} \\
0  & -\frac32 C_{12} &  -\frac32 C_{13}& 0 & \frac34 \left(\langle
n_f \rangle \
- 2D_f\right) + C_{15} & 0  \\
0  &  C_{13} &  C_{12} &  \frac12 C_{16} & 0 & D_f + C_{14}
\end{array}
\right) \label{e5}
\end{equation}
\end{widetext}

\section{The self-consistency cycle}
To close the self-consistency cycle, the following GF are needed
\begin{equation}
G_{12}=V \Gamma_0 (G_{22} + G_{23} ) \qquad G_{13}=V \Gamma_0
(G_{23} + G_{33} ) \; ,
\end{equation}
and
\begin{eqnarray}
G_{15} &=& V \Gamma_0 (G_{25} + G_{35} ) \nonumber \\
G_{25} &=& \frac{B_{+}}{F} G_{\pi 5} + \frac{(C_{5+}+x_{+}
C_{5-})}{F} \\
G_{35} &=& - \frac{B_{-}}{F} G_{\pi 5} + \frac{(C_{5-}+x_{-}
C_{5+})}{F} \nonumber
\end{eqnarray}
where we have in close analogy to (\ref{e16})
\begin{equation}
G_{\pi 5}= \frac{3}{2} \left( \frac{\langle n_f \rangle}{2} - D_f \right)
G_{11} + G_{55}^L
\end{equation}
and $C_{5+}=I_{25} \Gamma_{+}$, and $C_{5-}=I_{35} \Gamma_{-}$.

\bibliographystyle{apsrev}
\bibliography{Biblio}

\begin{thebibliography}{27}
\bibitem[{rec()}]{recipe}
\bibinfo{note}{We can choose: the higher order fields emerging from the
  equations of motion (i.e., the conservation of some spectral moments is
  assured)\cite{Mancini:98b}, the eigenoperators of some relevant interacting
  terms (i.e., the relevant interactions are correctly
  treated)\cite{Avella:02e}, the eigenoperators of the problem reduced to a
  small cluster\cite{Avella:03b}, the composite field describing the Kondo-like
  singlet emerging at low-energy in any electronic spin
  system\cite{Villani:00},...}
\end{thebibliography}

\end{document}